\documentclass[pre,twocolumn]{revtex4}

\usepackage{graphicx}
\usepackage{dcolumn}
\usepackage{amsmath}
\usepackage{amssymb}
\newcommand{\dotprod}{{\scriptscriptstyle \stackrel{\bullet}{{}}}}

\begin{document}

\title{Rayleigh-B\'{e}nard Convection with a Radial Ramp in Plate Separation}

\author{M. R. Paul}
 \email{mpaul@caltech.edu}
  \homepage{http://www.cmp.caltech.edu/~stchaos}
\author{M. C. Cross}
\affiliation{Department of Physics, California Institute of
Technology 114-36, Pasadena, California 91125}

\author{P. F. Fischer}
\affiliation{Mathematics and Computer Science Division, Argonne
National Laboratory, Argonne, Illinois 60439}

\date{\today}

\begin{abstract}
Pattern formation in Rayleigh-B\'{e}nard convection in a
large-aspect-ratio cylinder with a radial ramp in the plate
separation is studied analytically and numerically by performing
numerical simulations of the Boussinesq equations. A horizontal
mean flow and a vertical large scale counterflow are quantified
and used to understand the pattern wavenumber. Our results suggest
that the mean flow, generated by amplitude gradients, plays an
important role in the roll compression observed as the control
parameter is increased. Near threshold the mean flow has a
quadrupole dependence with a single vortex in each quadrant while
away from threshold the mean flow exhibits an octupole dependence
with a counter-rotating pair of vortices in each quadrant. This is
confirmed analytically using the amplitude equation and
Cross-Newell mean flow equation. By performing numerical
experiments the large scale counterflow is also found to aid in
the roll compression away from threshold but to a much lesser
degree. Our results yield an understanding of the pattern
wavenumbers observed in experiment away from threshold and suggest
that near threshold the mean flow and large scale counterflow are
not responsible for the observed shift to smaller than critical
wavenumbers.
\end{abstract}

\pacs{47.54.+r,47.20.Bp,47.27.Te}

\maketitle

\section{Introduction}
\label{section:introduction} Rayleigh-B\'{e}nard convection in a
thin horizontal fluid layer heated from below is a canonical
example of pattern formation in a continuous dissipative system
far from equilibrium~\cite{cross:1993}. Under various conditions,
wavenumber selection mechanisms have been
identified~\cite{cross:1986,catton:1988,cross:1993,getling:1998}
that reduce the band of stable wavenumbers, sometimes to a single
value. One such selection mechanism occurs when there is a
one-dimensional spatial variation or ramping of the control
parameter $\epsilon$; where $\epsilon \equiv (R-R_c)/R_c$, $R$ is
the Rayleigh number, and $R_c$ its critical value
~\cite{eagles:1980,kramer:1982,pomeau:1983,kramer:1985,buell:1986:ramp,bajaj:1999}.
This can be accomplished by varying the plate separation $d$ such
that $\epsilon$ goes from $\epsilon=\epsilon_o>0$ in the bulk of
the layer (i.e., the unramped region) to $\epsilon<0$ as a lateral
boundary is approached. It is expected that in the idealized case
of an infinitely gradual one-dimensional ramp the wavenumber will
equal $k_c$~ at the position where the layer depth yields $R_c$.
It has been shown under very general conditions, that as long as
the layer becomes critical somewhere along the ramp this is
sufficient to fix the wavenumber $k_s$ in the bulk and over the
rest of the ramp~\cite{kramer:1982}. For slightly supercritical
conditions it is expected that the selected wavenumber in the bulk
can be expressed as
\begin{equation}
k_s=\tilde{k}_c+\alpha\epsilon_o
\end{equation}
where $\tilde{k}_c=k_c=3.117$~\cite{bajaj:1999} and $\alpha$
depends on the Prandtl number, $\sigma$, and the specifics of the
ramp.

\begin{figure}[tbh]
\begin{center}
\includegraphics[width=3.0in]{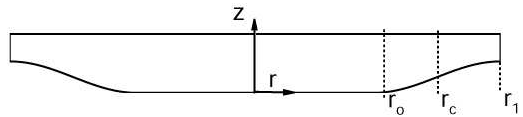}
\end{center}
\caption{A vertical cross-section of a cylindrical convection
layer with a radial ramp in plate separation, $r_0$ defines where
the ramp begins, $r_1$ defines where the ramp ends and $r_c$ is
where the plate separation corresponds to the critical Rayleigh
number. The ramp shown is a cosine ramp given by
Eq.~(\ref{eq:platesep}). For presentation purposes we show a steep
ramp with $\delta_r=0.25$ (this domain is not used in the
simulations).} \label{fig:config}
\end{figure}

Recent Rayleigh-B\'{e}nard convection
experiments~\cite{bajaj:1999,ahlers:2001} in a cylindrical cell
with a two-dimensional radial ramp in plate separation have
generated intriguing results. Specifically, the plate separation
as a function of radius used in experiment is, using the layer
depth $d$ to nondimensionalize,
\begin{equation}
 h(r) = \left\{ \begin{array}{ll}
   1, & \mbox{$r < r_0$} \\
   1 - {\delta_r} \left[ 1- \cos \left( \frac{r-r_0} {r_1-r_0} \pi \right) \right], & \mbox{$r \ge
   r_0$}
\end{array}\right.
\label{eq:platesep}
\end{equation}
where $\delta_r$ is a constant and the radius values $r_0$ and
$r_1$ are the locations where the ramp begins and ends,
respectively. The ramp always extends to the sidewall. A schematic
of the cosine ramp is shown in Fig.~\ref{fig:config}; note that
$r_0$ and $r_1$ are geometric constants but $r_c$, the location
where the plate separation yields $R_c$, is a function of the ramp
shape and $\epsilon_0$ such that, for a ramp given by
Eq.~(\ref{eq:platesep}), $\partial r_c / \partial \epsilon_0 > 0$.
%%%%%%%%%%%%%%%%%%%%%%%%%%%%%%%%%%%%%%%%%%%%%%%%%%%%%%%%%%%%%%%%%%%%%%%%%%%%%
%%%%%%%%%%%%%%%%%%%%%%%%%%%%%%%%%%%%%%%%%%%%%%%%%%%%%%%%%%%%%%%%%%%%%%%%%%%%%

Experiments using the cosine ramp defined by
Eq.~(\ref{eq:platesep}) have yielded unexpected results for the
wavenumber~\cite{ahlers:2001,bajaj:1999}. Mean pattern wavenumber
measurements (using the Fourier methods discussed in
\cite{morris:1993}) yielded $\tilde{k}_c =2.97 < k_c$.
Additionally, measurements of the local wavenumber defined at each
position in space (method discussed in \cite{egolf:1998})
displayed interesting variation as $\epsilon_0$ is increased. For
the time independent patterns the bulk region, $r \le r_0$,
contains approximately straight parallel rolls. Near threshold,
$\epsilon_0 \lesssim 0.048$, a centered egg-shaped domain of
convection rolls with small wavenumber, $k < k_c$, extends through
the convection cell with the long-axis \textit{parallel} to the
roll axes. A dramatic roll expansion from $k \approx 3.6$ at the
edge where the ramp begins to $k \approx 2.6$ in the center of the
domain is observed. As $\epsilon_0$ increases, the wavenumber
field evolves into a domain characterized by large wavenumbers
extending through the layer with the long-axis
\textit{perpendicular} to the roll axes (see Fig.~3 of
\cite{bajaj:1999}). Similar experiments without a ramp do not
exhibit this wavenumber
behavior~\cite{hu:1993,hu:1994:prl,hu:1995}.

In the experiments with a radial ramp in plate separation, for
$\epsilon_0 \lesssim 0.03$, time dependent states were found
through the repeated formation of defects via an Eckhaus mechanism
consistent with the local wavenumbers crossing the Eckhaus
stability boundary for an ideal infinite layer of two-dimensional
rolls. Also in the ramped experiments, for $\epsilon_0 \gtrsim
0.18$, defects were formed via a skewed varicose mechanism
consistent with the local wavenumber exceeding the skewed varicose
stability boundary for an ideal infinite layer of two-dimensional
rolls similar to what has been observed in experiments in unramped
cylindrical domains with rigid sidewalls~\cite{croquette:1989}.

It has been suggested that these features may be the result of the
interaction of the convective roll pattern and weak large scale
flows~\cite{bajaj:1999}. The visualization and quantification of
these large scale flows is not possible in the current generation
of experiments. However, we are able to make these measurements by
performing full numerical simulations with a new spectral element
code (discussed further in Section~\ref{section:numerical
simulation}). We utilize the complete knowledge of the flow field
together with analytical results valid near threshold to explore
this further.

\section{Large Scale Flows}
\label{section:lsf} In this work the terminology large scale flows
is used to describe flows that extend over distances larger than
that of the convection roll scale. We would like to distinguish
between two different large scale flows: {\em large scale
counterflow} and {\em mean flow}.

\subsection{Large Scale Counterflow}
\begin{figure}[tbh]
\begin{center}
\includegraphics[width=3.0in]{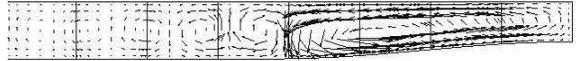}
\end{center}
\caption{Velocity vectors illustrating the large scale counterflow
for a vertical slice of a cylindrical convection layer for
subcritical conditions, $\epsilon_0=-0.063$. The ramp parameters
are $r_0=7.66$, $r_1=10$, and $\delta_r=0.15$. A steep ramp is
shown to clearly illustrate the flow. Solid vertical lines
indicate the boundaries of the spectral element grid used in the
simulation, only a portion of the layer is shown emphasizing the
ramped region.} \label{fig:yslice}
\end{figure}

In the presence of a spatial ramp in plate separation a large
scale counterflow is present for all values of the bulk control
parameter $\epsilon_0$, including $\epsilon_0<0$. Warm fluid
ascends the ramp eventually reaching the sidewall and is forced to
flow back toward the center of the domain over the cold top wall
causing it to descend resulting in a large zone of circulation in
the vertical plane over the ramp. The magnitude of the large scale
counterflow depends upon the specifics of the ramp and is roughly
independent of $\epsilon_0$ and $\sigma$~\cite{walton:1982}.
Figure~\ref{fig:yslice} illustrates this with a vertical slice
from a three-dimensional numerical simulation where the entire
convection layer is subcritical. As shown for this subcritical
case the fluid motion of the large scale counterflow generates
axisymmetric convection near the base of the ramp which extends a
couple of roll widths toward the center of the domain. For the
gradual ramp used in experiment the large scale counterflow is
small in magnitude and has not been measured.
%%%%%%%%%%%%%%%%%%%%%%%%%%%%%%%%%%%%%%%%%%%%%%%%%%%%%%%%%%%%%%%%%%%%%%%%%%%%%
%The magnitude of the flow velocity is $|\vec{u}|\sim 0.4$.
%%%%%%%%%%%%%%%%%%%%%%%%%%%%%%%%%%%%%%%%%%%%%%%%%%%%%%%%%%%%%%%%%%%%%%%%%%%%%

\subsection{Mean Flow}
In discussing the mean flow it will be convenient to first present
the governing heat and fluid equations. The velocity $\vec{u}$,
temperature $T$, and pressure $p$, evolve according to the
Boussinesq equations,
\begin{eqnarray}
  {\sigma}^{-1} \left(
  {\partial}_t + \vec{u} \dotprod \vec{\nabla} \right) \vec{u}
  &=&  -\vec{\nabla} p + RT \hat{z} + \nabla^2 \vec{u}  , \label{eq:mom}\\
  \left( {\partial}_t + \vec{u} \dotprod \vec{\nabla} \right) T
  &=& \nabla^2 T  , \label{eq:energy}\\
  \vec{\nabla} \dotprod \vec{u} &=& 0,
  \label{eq:mass}
\end{eqnarray}
where $\partial_t$ indicates time differentiation, and $\hat{z}$
is a unit vector in the vertical direction opposite to gravity.
The equations are nondimensionalized in the standard manner using
the layer depth $d$, the vertical diffusion time for heat
${\tau}_v \equiv d^2/\kappa$ where $\kappa$ is the thermal
diffusivity, and $\Delta T$ the temperature difference between the
top and bottom surfaces, as the length, time, and temperature
scales, respectively. The lower and upper surfaces are no-slip and
are held at constant temperature. The sidewalls are also no-slip
and, unless otherwise noted, are insulating.

The mean flow field, $\vec{U}(x,y)$, is the horizontal velocity
integrated over the depth and originates from the Reynolds stress
induced by pattern distortions. As illustrated by the fluid
equations, Eqs.~(\ref{eq:mom}) and~(\ref{eq:mass}), it is evident
that the pressure is not an independent dynamic variable. The
pressure is determined implicitly to enforce incompressibility,
\begin{equation} \label{eq:pressure}
\nabla^2 p = -\sigma^{-1} \vec{\nabla} \dotprod \left[ \left(
\vec{u} \dotprod \vec{\nabla} \right) \vec{u} \right] + R
\partial_z T.
\end{equation}
Focussing on the nonlinear Reynolds stress term and rewriting the
pressure as $p = p_o(x,y) + \bar{p}(x,y,z)$ yields,
\begin{equation} \label{eq:pressure_p0}
p_o(x,y) \sim \sigma^{-1} \int dx' dy' \ln \left( 1 / \left| r-r'
\right| \right) \left< \vec{\nabla}' \dotprod \left[ \left(
\vec{u} \dotprod \vec{\nabla} \right) \vec{u} \right] \right>_z ,
\end{equation}
where $\left< \cdot \right>_z$ represents an average in the z
direction. In Eq.~(\ref{eq:pressure_p0}) the $\ln(1/|r-r'|)$ is
not exact, in order to be more precise the finite system Green's
function would be required; however the long range behavior
persists. This gives a contribution to the pressure that depends
on distant parts of the convection pattern. The Poiseuille-like
flow driven by this pressure field subtracts from the Reynolds
stress induced flow leading to a divergence free horizontal flow
that can be described in terms of a vertical vorticity.

Near threshold an explicit expression for the mean flow
is~\cite{cross:1984}
\begin{equation}
\vec{U}(x,y) = - \gamma \vec{k} \vec{\nabla}_{\perp} \dotprod
\left( \vec{k} A^2 \right) - \vec{\nabla}_{\perp}p_o(x,y)
\label{eq:meanflow}
\end{equation}
where $\gamma$ is a coupling constant given by $\gamma = 0.42
{\sigma}^{-1} (\sigma+0.34)(\sigma+0.51)^{-1}$, $A^2$ is the
convection amplitude normalized so that the convective heat flow
per unit area relative to the conducted heat flow at $R_c$ is
$|A|^2R/R_c$, $p_o$ is a slowly varying pressure, see
Eq.~(\ref{eq:pressure_p0}), and $\vec{\nabla}_{\perp}$ is the
horizontal gradient operator
[see~\cite{siggia:1981,newell:1990:jfm} for the complete analysis
and more details]. The mean flow is important not because of its
strength; under most conditions the magnitude of the mean flow is
substantially smaller than the magnitude of the roll flow making
it extremely difficult to quantify experimentally. The mean flow
is important because it is a nonlocal effect acting over large
distances (many roll widths) and changes important general
predictions of the phase equation~\cite{cross:1984}. The mean flow
is driven by roll curvature, roll compression and gradients in the
convection amplitude. The resulting mean flow advects the pattern
giving an additional slow time dependence. It is important to
note, that unlike the long range counterflow, the magnitude of the
mean flow vanishes when the convection layer becomes critical,
$|\vec{U}| \sim \epsilon_0$ for $\epsilon_0 \ll 1$.

\section{Numerical Simulation}
\label{section:numerical simulation} We have performed full
numerical simulations of the governing fluid and heat equations,
Eqs.~(\ref{eq:mom})-(\ref{eq:mass}), in a cylindrical geometry
with a radial ramp in plate separation using a parallel spectral
element algorithm (described in detail elsewhere
\cite{fischer:1997}, see \cite{paul:2001,{paul:2002:physd}} for
related applications).

\begin{figure}[tbh]
\begin{center}
\includegraphics[width=2.0in]{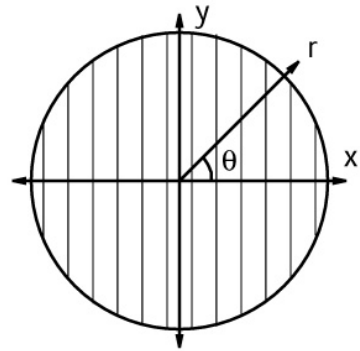}
\end{center}
\caption{Cartesian~$(x,y)$ and polar~$(r,\theta)$ coordinates
defined on a mid-depth horizontal cross-section of a cylindrical
convection layer containing a field of x-rolls described by
$\vec{k}=k_o \hat{x}$.} \label{fig:config_xy}
\end{figure}

For discussion purposes it will be convenient to define
Cartesian~$(x,y)$ and polar~$(r,\theta)$ coordinates centered on a
mid-depth horizontal slice of a cylindrical convection layer
containing a field of straight parallel x-rolls with wavevector
$\vec{k}=k_o \hat{x}$ as shown in Fig.~\ref{fig:config_xy}. The
x-axis is perpendicular to the roll axes, the y-axis is parallel
to the roll axes and $\theta$ measures the angle from the positive
x-axis.
%%%%%%%%%%%%%%%%%%%%%%%%%%%%%%%%%%%%%%%%%%%%%%%%%%%%%%%%%%%%%%%%%%%%%%%%%%%%%
%%%%%%%%%%%%%%%%%%%%%%%%%%%%%%%%%%%%%%%%%%%%%%%%%%%%%%%%%%%%%%%%%%%%%%%%%%%%%

We have investigated the results found in experiment by performing
simulations on a ramped cylindrical convection layer for a variety
of scenarios and initial conditions. Simulations were performed
over the range $\epsilon_0 \lesssim 0.2$ and for simulation times
of $t_f > \tau_h$ where $\tau_h \equiv r_0^2$ is the time required
for heat to diffuse horizontally across the bulk region of the
layer which has been suggested as the earliest time scale for the
flow field to reach equilibrium~\cite{cross:1984}.

The mean flow present in the simulation flow fields,
$\vec{U}_s(x,y)$, is investigated by calculating the depth
averaged horizontal velocity,
\begin{equation}
\vec{U}_s(x,y)=\int^1_0 \vec{u}_{\perp}dz
\end{equation}
where $\vec{u}_{\perp}$ is the horizontal velocity field.
Furthermore it will be convenient to work with the vorticity
potential, $\zeta$, defined as
\begin{equation} \label{eq:vorticity_potential}
\nabla^2_\perp \zeta = -\hat{z} \dotprod \left( \nabla_{\perp}
\times \vec{U}_s \right) = - \omega_z
\end{equation}
where $\omega_z$ is the vertical vorticity and $\nabla^2_\perp$ is
the horizontal Laplacian.
%%%%%%%%%%%%%%%%%%%%%%%%%%%%%%%%%%%%%%%%%%%%%%%%%%%%%%%%%%%%%%%%%%%%%%%%%%%%%
\section{Analytical Development}
\label{section:analytical development} Near threshold, assuming
straight parallel rolls, it is possible to approximately determine
$\omega_z$ analytically. It will be convenient to start from
$\omega_z$ given by the vertical component of the curl of
Eq.~(\ref{eq:meanflow}),
\begin{equation}
{\omega}_z = \hat{z} \dotprod \left( \vec{\nabla}_{\perp} \times
\vec{U} \right) = - \gamma \hat{z} \dotprod \vec{\nabla}_{\perp}
\times \left[ \vec{k} \vec{\nabla}_{\perp} \dotprod \left( \vec{k}
|A|^2 \right) \right]. \label{eq:wz_general}
\end{equation}
Consider a cylindrical convection layer with a radial ramp in
plate separation containing a field of x-rolls given by
$\vec{k}=k_o \hat{x}$. The amplitude can be represented for large
$\epsilon_0$, using an adiabatic approximation, as
$|A|^2=\epsilon(r)/g_o$ for $\epsilon>0$ and $|A|^2=0$ for
$\epsilon(r)<0$, making the amplitude a function of radius only
$|A|^2=f(r)$. This approximation is good except for the kink at
$r_c$ where $\epsilon=0$. Inserting $|A|^2=f(r)$ into
Eq.~(\ref{eq:wz_general}) yields, after some manipulation, the
following expression for the vertical vorticity,
\begin{equation}
{\omega}_z = \frac{\gamma {k_o}^2}{2} \left[ \frac{d^2|A|^2}{dr^2}
- \frac{1}{r} \frac{d|A|^2}{dr} \right] \sin 2 \theta.
\label{eq:wz}
\end{equation}
To correct for nonadiabaticity and to smooth $|A(r)|^2$ near
$r_c$, the one-dimensional time independent amplitude equation is
solved,
\begin{equation}
0 = \epsilon (r) A + {{\xi}_o}^2 {\cos}^2 {\theta}
\frac{\partial^2A}{\partial r^2} - g_o |A|^2A, \label{eq:amp_dim}
\end{equation}
where angular derivatives have been assumed small, ${{\xi}_o}^2 =
0.148$, $g_o=0.6995-0.0047\sigma^{-1}+0.0083\sigma^{-2}$ and
$\epsilon(r)$ is determined by
\begin{equation}
 \epsilon(r) = \left\{ \begin{array}{ll}
   \epsilon_0, & \mbox{$r < r_0$} \\
   \epsilon_0 (h^3-h_c^3)/(1-h_c^3), & \mbox{$r \ge
   r_0$}
\end{array}\right.
\label{eq:eps}
\end{equation}
where $h_c=h(r_c)$. In Eq.~(\ref{eq:wz_general}) $\omega_z$ is
dominated by radial derivatives so Eq.~(\ref{eq:wz}) is still a
good approximation for the case $|A|^2=f(r,\theta)$, $d/dr
\rightarrow
\partial/\partial r$. Due to the angular dependence of the
nonadiabaticity $|A|^2$ is now $\theta$ dependent  which will
induce higher angular harmonics in $\omega_z$. We will neglect
these higher harmonics, assume a $\sin 2 \theta$ dependence and
approximately evaluate the magnitude using Eqs.~(\ref{eq:wz})
and~(\ref{eq:amp_dim}) at $\theta=\pi/4$.
Equation~(\ref{eq:amp_dim}) can be rewritten in a more convenient
form as,
\begin{equation}
0 = \frac{\epsilon(\bar{r})}{\epsilon_0} \bar{A} +
\frac{\partial^2\bar{A}}{\partial \bar{r}^2} - \bar{A}^3
\label{eq:amp_nondim}
\end{equation}
where $r = \left( {{\epsilon}_0}^{-1/2} {\xi}_o \cos \theta
\right) \bar{r}$ and $A = {(\epsilon_0/g_o)}^{1/2}
\bar{A}$~\cite{cross:1984} (since the amplitude goes to zero it
can be shown that $A$ is now a real
quantity~\cite{cross:1983:cdhs}). Equation~(\ref{eq:amp_nondim})
is solved numerically using the boundary conditions
$\partial_{\bar{r}} \bar{A}=0$ at $\bar{r} = 0$, and $\bar{A}=0$
at $\bar{r} = {\bar{r}}_1$.

These analytical results are now used to investigate the vertical
vorticity generation in a large radially ramped cylindrical
convection layer. We start by looking at the configuration used in
the recent experiments with input parameters: $r_0=42.29$,
$r_1=101.33$, ${\delta}_r=0.036$ and $\sigma=0.87$.

\begin{figure}[tbh]
\begin{center}
\includegraphics[width=3.0in]{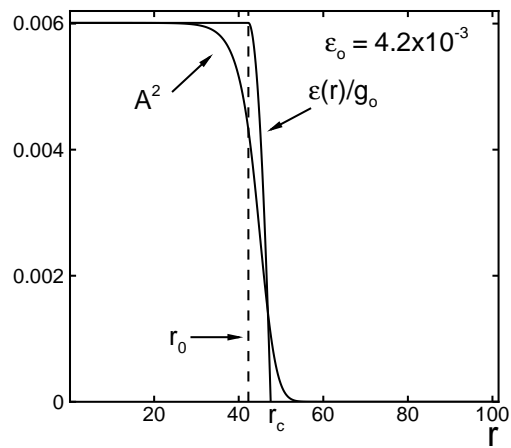}
\end{center}
\caption{The solution of Eq.~(\ref{eq:amp_nondim}) plotted as
$A^2(r)$ for $r_0=42.29$, $r_c=47.56$, $r_1=101.33$,
$\delta_r=0.036$, $\sigma=0.87$ and $\epsilon_0 = 4.20 \times
10^{-3}$. Also shown for comparison is $\epsilon(r)/g_o$.}
\label{fig:amp_eps_g101_1715}
\end{figure}

\begin{figure}[tbh]
\begin{center}
\includegraphics[width=3.0in]{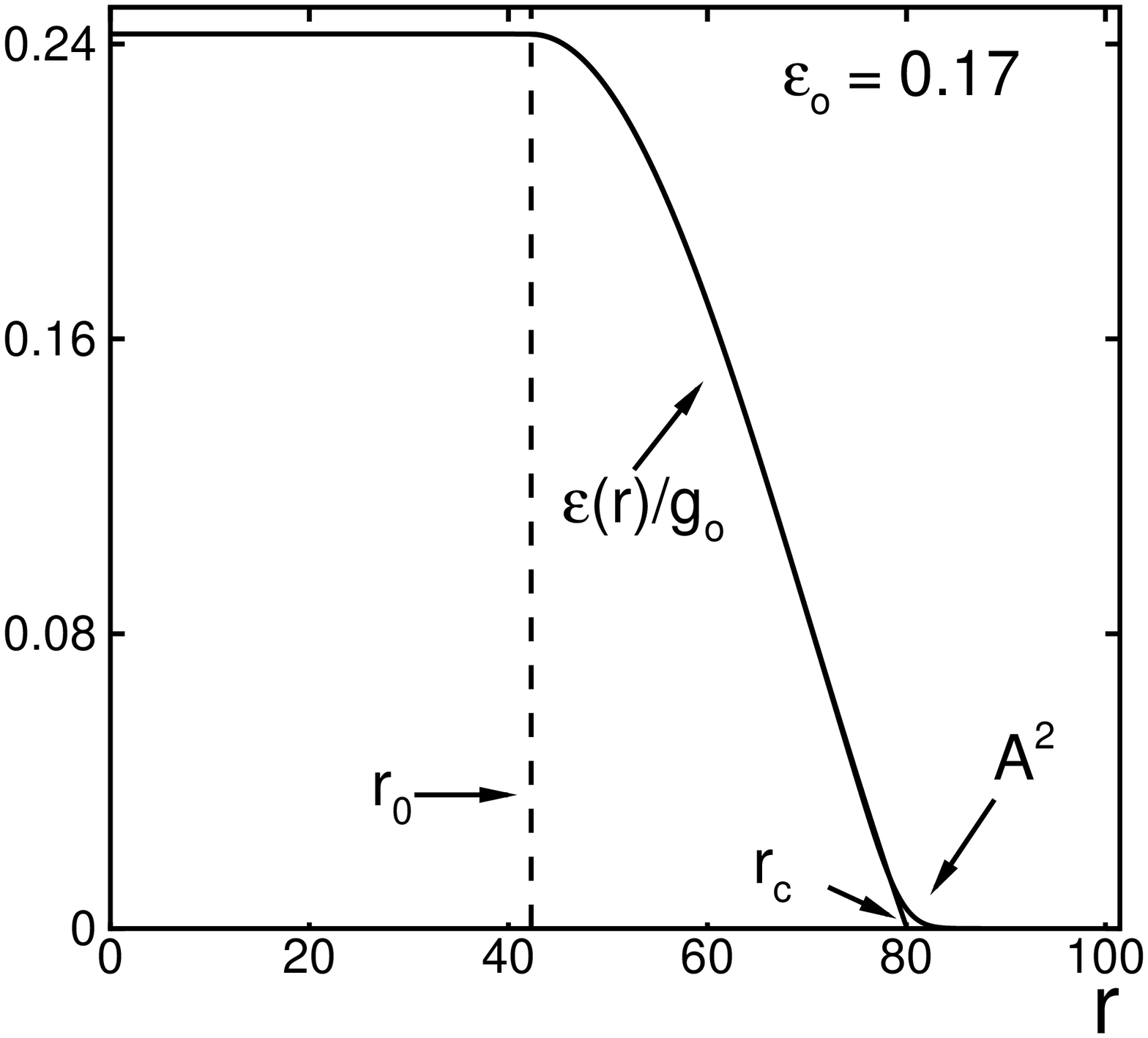}
\end{center}
\caption{The solution of Eq.~(\ref{eq:amp_nondim}) plotted as
$A^2(r)$ for $r_0=42.29$, $r_c=80.05$, $r_1=101.33$,
$\delta_r=0.036$, $\sigma=0.87$ and $\epsilon_0 = 0.17$. Also
shown for comparison is $\epsilon(r)/g_o$.}
\label{fig:amp_eps_g101_2000}
\end{figure}

\begin{figure}[tbh]
\begin{center}
\includegraphics[width=3.0in]{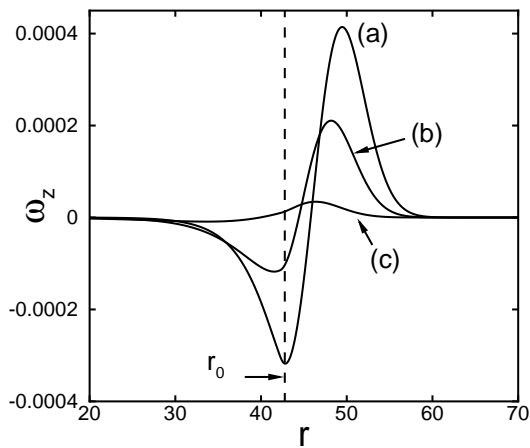}
\end{center}
\caption{Dependence of the radial variation of
$\omega_z=\omega_z(r)$ with $\epsilon_0$ as determined
analytically from Eqs.~(\ref{eq:wz}) and (\ref{eq:amp_nondim})
illustrating the evolution from a mean flow with a quadrupole
dependence to an octupole dependence as $\epsilon_0$ increases.
Shown explicitly are $\omega_z$ curves for $\epsilon_0 = 7.2
\times 10^{-3}$, $4.2 \times 10^{-3}$, and $1.3 \times 10^{-3}$
with the parameters $r_0=42.29$, $r_1=101.33$, $\delta_r=0.036$
and $\sigma=0.87$ labelled (a)-(c), respectively.}
\label{fig:vort_var_sm_g101}
\end{figure}

When ${\epsilon_0}^{1/2} {\xi_o}^{-1} (r_c-r_0) \lesssim 1$ the
amplitude $A^2(r)$ is unable to adiabatically follow the ramp,
this nonadiabaticity results in a considerable deviation from
$\epsilon(r)/g_o$ as shown in Fig.~\ref{fig:amp_eps_g101_1715}.
However, when ${\epsilon_0}^{1/2} {\xi_o}^{-1} (r_c-r_0) \gg 1$
the amplitude $A^2(r)$ follows $\epsilon(r)/g_o$ adiabatically
almost over the entire ramp except for the small kink at $r_c$ as
shown in Fig.~\ref{fig:amp_eps_g101_2000}. The structure of
$\omega_z$ depends upon this adiabaticity and is shown for various
values of $\epsilon_0$ in Fig.~\ref{fig:vort_var_sm_g101} where
the $\sin 2\theta$ dependence has been removed by choosing
$\theta=\pi/4$.

If the $\theta$ dependence is included it is evident from
Fig.~\ref{fig:vort_var_sm_g101} that the vertical vorticity has a
quadrupole angular structure for small $\epsilon_0$, i.e. four
lobes of alternating positive and negative vorticity with one lobe
per quadrant, and makes a transition to an octupole angular
dependence for larger $\epsilon_0$, octupole in the sense of an
inner, $r \lesssim r_c$, and outer, $r \gtrsim r_c$, quadrupole.
In addition, since $\partial r_c/ \partial \epsilon_0
> 0$ there is a radial shift of the vorticity curves as
$\epsilon_0$ is increased.

In all cases the amplitude $A^2(r)$ decreases monotonically with
$r$ and as a result $-r^{-1}~d|A|^2/dr \ge0$ thus generating only
positive vorticity. However, the term $d^2|A|^2/dr^2$ can be of
either sign and is responsible for the quadrupole and octupole
angular structure in $\omega_z$.
%%%%%%%%%%%%%%%%%%%%%%%%%%%%%%%%%%%%%%%%%%%%%%%%%%%%%%%%%%%%%%%%%%%%%%%%%%%%%
%%%%%%%%%%%%%%%%%%%%%%%%%%%%%%%%%%%%%%%%%%%%%%%%%%%%%%%%%%%%%%%%%%%%%%%%%%%%%
%%%%%%%%%%%%%%%%%%%%%%%%%%%%%%%%%%%%%%%%%%%%%%%%%%%%%%%%%%%%%%%%%%%%%%%%%%%%%
%%%%%%%%%%%%%%%%%%%%%%%%%%%%%%%%%%%%%%%%%%%%%%%%%%%%%%%%%%%%%%%%%%%%%%%%%%%%%
%%%%%%%%%%%%%%%%%%%%%%%%%%%%%%%%%%%%%%%%%%%%%%%%%%%%%%%%%%%%%%%%%%%%%%%%%%%%%
%%%%%%%%%%%%%%%%%%%%%%%%%%%%%%%%%%%%%%%%%%%%%%%%%%%%%%%%%%%%%%%%%%%%%%%%%%%%%
As $\epsilon_0$ approaches zero the nonadiabaticity of $A^2(r)$
increases until $A^2(r)$ exhibits a quadratic fall-off with $r$
for $r \ll r_0$ resulting in $\omega_z(r) \ge 0$ for all $r$.

\begin{figure}[tbh]
\begin{center}
\includegraphics[width=3.0in]{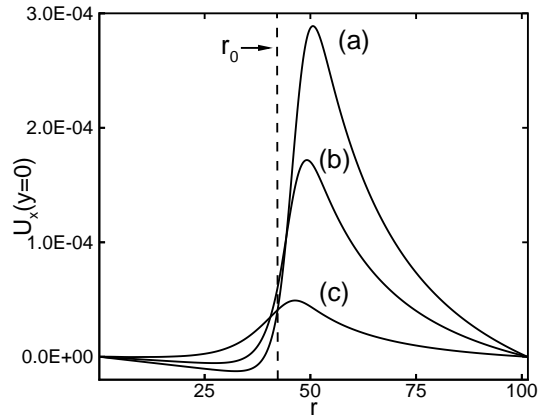}
\end{center}
\caption{Variation of the mean flow, $U_x(y=0)$ with $\epsilon_0$
as determined from Eq.~(\ref{eq:vorticity_potential}). Shown
explicitly are curves for $\epsilon_0 = 7.2 \times 10^{-3}$, $4.2
\times 10^{-3}$, and $1.3 \times 10^{-3}$ labelled (a)-(c),
respectively with the parameters $r_0=42.29$, $r_1=101.33$,
$\delta_r=0.036$ and $\sigma=0.87$.} \label{fig:mf_var_sm_g101}
\end{figure}

The mean flow generated by these vorticity distributions is
determined by solving Eq.~(\ref{eq:wz_general}) with the boundary
condition $\zeta(r_1)=0$. The vorticity potential is related to
the mean flow in polar coordinates by $(U_r,U_\theta)=(r^{-1}
\partial_\theta \zeta, -\partial_r \zeta)$. The vorticity
potential is expanded radially in second order Bessel functions
while maintaining the $\sin 2 \theta$ angular dependence. Of
particular interest is the mean flow perpendicular to the
convection rolls, $U_r(\theta=0)$ or equivalently $U_x(y=0)$,
which is shown in Fig.~\ref{fig:mf_var_sm_g101}.
%%%%%%%%%%%%%%%%%%%%%%%%%%%%%%%%%%%%%%%%%%%%%%%%%%%%%%%%%%%%%%%%%%%%%%%%%%%%%
%%%%%%%%%%%%%%%%%%%%%%%%%%%%%%%%%%%%%%%%%%%%%%%%%%%%%%%%%%%%%%%%%%%%%%%%%%%%%

As expected, regions of negative and positive vorticity yield
corresponding negative and positive values of the mean flow. As
$\epsilon_0$ vanishes $U \ge 0$ for all $r$ providing a mechanism
for roll expansion in the bulk. For larger $\epsilon_0$ the mean
flow becomes larger in magnitude and increasingly negative for $r
\le r_0$ providing a mechanism for roll compression.

To make the connection between mean flow and wavenumber
quantitative it is noted that the wavenumber variation resulting
from a mean flow across a field on x-rolls can be determined from
the one-dimensional phase equation,
\begin{equation} \label{eq:phase}
U \partial_x \phi = D_\parallel \partial_{xx}\phi
\end{equation}
where the wavenumber is the gradient of the phase,
$k=\partial_x\phi$, $D_\parallel=\xi_o^2 \tau_o^{-1}$, and
$\tau_o^{-1}=19.65 \sigma (\sigma+0.5117)^{-1}$~\cite{cross:1993}.
Assuming that for $\epsilon_0 \ll 1$ the wavenumber is
approximately $k \approx k_c$ everywhere and that the rolls are
exposed to a constant mean flow the wavenumber change over the
bulk can be expressed as
\begin{equation} \label{eq:delta_k}
\Delta k = k(r_0) - k(r=0) = U k_c D_\parallel^{-1} r_0.
\end{equation}
For example, for curve (c) in Fig.~\ref{fig:mf_var_sm_g101} the
maximum value of the mean flow is $U=4.92 \times 10^{-5}$ which
yields a small roll expansion of $\Delta k = 0.0035$. If the mean
flow were solely responsible for the dramatic roll expansion seen
in experiment of $\Delta k \approx 1.0$ ($\epsilon_0=0.012$ see
Fig.~3(a) of \cite{bajaj:1999}) a mean flow of $U \approx 0.014$
would be required, which is not found in the analytic results.
%%%%%%%%%%%%%%%%%%%%%%%%%%%%%%%%%%%%%%%%%%%%%%%%%%%%%%%%%%%%%%%%%%%%%%%%%%%%%
\section{Discussion}
\label{section:discussion} The large scale flows discussed in
Section~\ref{section:lsf} cannot be measured in current
experiments placing us in a unique position to use the complete
flow field information from our full numerical simulations of
Eqs.~(\ref{eq:mom})-(\ref{eq:mass}) together with the analytical
results of Section~\ref{section:analytical development} to
investigate how the mean flow and the large scale counterflow
induce wavenumber distortions and the variation of this distortion
with Rayleigh number.

It is computationally expensive to perform full three-dimensional
numerical simulations for the very large system used in
experiment. We have, however, performed a variety of simulations
for radially ramped cylindrical convection layers. The full
three-dimensional simulations are of smaller spatial extent with
the precise ramp defined by Eq.~(\ref{eq:platesep}) and the
specific input parameters: $r_0=11.31$, $r_1=20$,
$\delta_r=0.036$, and $\sigma=0.87$. Two-dimensional simulations
of a vertical slice of a three-dimensional domain (see
Fig.~\ref{fig:config}) were also performed for both the large
experimental configuration and the smaller computational domain
just described. Three-dimensional simulations were also conducted
without a large scale counterflow by a specific choice of ramp
parameters that will be discussed below.

\begin{figure}[ht]
\begin{center}
\includegraphics[width=3.0in]{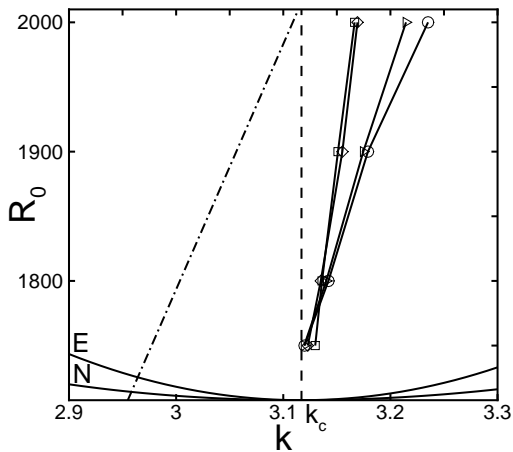}
\end{center}
\caption{Comparison of the mean wavenumber variation as a function
of the bulk Rayleigh number (i.e. for $r \le r_0$), $R_0$, between
simulation (solid lines with symbols) and experiment
(dashed-dotted line)~\cite{bajaj:1999}. Unless otherwise noted
$r_0=11.31$, $r_1=20$, $\delta_r=0.036$, and $\sigma=0.87$. The
symbols represent: $(\circ)$ three-dimensional simulations,
($\square$) two-dimensional simulations, ($\triangleright$)
three-dimensional simulations for a specific ramp construction
without a large scale counterflow, and ($\diamond$)
two-dimensional simulations with $r_0=42.29$, and $r_1=101.33$.
Dark solid lines denote the approximate location of the neutral
(N) and Eckhaus (E) stability boundaries for an ideal infinite
layer of parallel rolls.} \label{fig:wnall}
\end{figure}

Initially we consider prescribed x-roll initial conditions given
by $\vec{k}=k_c\hat{x}$. Other initial conditions such as random
thermal perturbations or initial x-rolls of varying wavenumbers
were also investigated and found not to affect the final pattern
wavenumber or any of the conclusions drawn. Simulations were
performed for $\epsilon_0 = 0.025$, 0.054, 0.113, and 0.171.
Figure~\ref{fig:wnall} compares the wavenumbers found in these
simulations with recent experiments and will be discussed in
detail below.
%%%%%%%%%%%%%%%%%%%%%%%%%%%%%%%%%%%%%%%%%%%%%%%%%%%%%%%%%%%%%%%%%%%%%%%%%%%%%
%%%%%%%%%%%%%%%%%%%%%%%%%%%%%%%%%%%%%%%%%%%%%%%%%%%%%%%%%%%%%%%%%%%%%%%%%%%%%

\begin{figure}[ht]
\begin{center}
\includegraphics[width=2.5in]{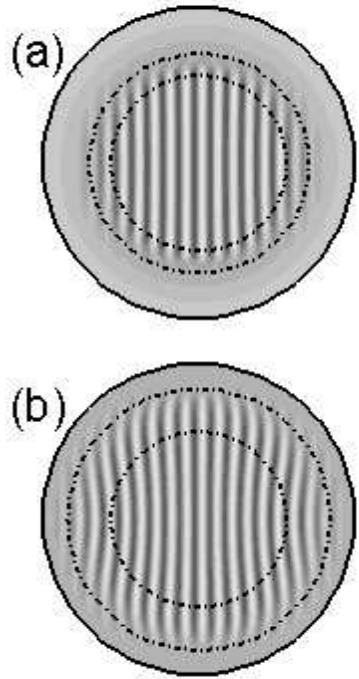}
\end{center}
\caption{Final convection patterns for $\epsilon_0 = 0.025$ and
$0.17$ are shown in panels (a) and (b), respectively. Shaded
contours of the thermal perturbation are shown with dark regions
representing cool descending fluid and light regions warm
ascending fluid. The inner dotted circle indicates where the ramp
begins, $r_0$, and the outer dotted circle indicates where the
convection layer becomes critical, $r_c$. Simulation parameters,
$r_0=11.31$, $r_1=20$, $\delta_r=0.036$, $\sigma=0.87$.}
\label{fig:tp}
\end{figure}

The final patterns in the simulations maintain the x-roll
configuration imposed by the initial conditions.
Figure~\ref{fig:tp} displays the final pattern observed for
three-dimensional simulations with $\epsilon_0=0.025$ in panel (a)
and $\epsilon_0=0.171$ in panel (b). Figure~\ref{fig:tp}a
illustrates that near threshold the convection rolls exhibit very
little curvature indicating that the assumption of straight
parallel x-rolls in Section~\ref{section:analytical development}
is valid. There is more roll curvature apparent in
Fig.~\ref{fig:tp}b as would be expected for larger $\epsilon_0$.
Figure~\ref{fig:tp} also illustrates the decreasing size of the
subcritical region as the supercriticality of the bulk increases.
All simulations settled to a time independent state.
%%%%%%%%%%%%%%%%%%%%%%%%%%%%%%%%%%%%%%%%%%%%%%%%%%%%%%%%%%%%%%%%%%%%%%%%%%%%%
%%%%%%%%%%%%%%%%%%%%%%%%%%%%%%%%%%%%%%%%%%%%%%%%%%%%%%%%%%%%%%%%%%%%%%%%%%%%%

\begin{figure}[ht]
\begin{center}
\includegraphics[width=2.5in]{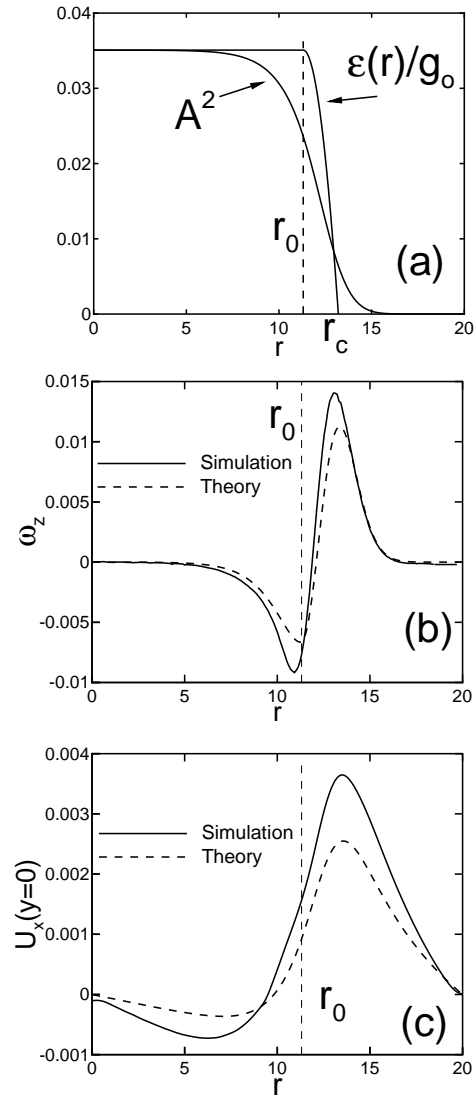}
\end{center}
\caption{Panel~(a) shows the solution of Eq.~(\ref{eq:amp_nondim})
plotted as $A^2(r)$, shown for comparison is $\epsilon(r)/g_o$.
Panel~(b) compares the vertical vorticity found analytically from
Eq.~(\ref{eq:amp_nondim}) with an angular average, weighted by
$\sin 2 \theta$, of the vertical vorticity from simulation.
Panel~(c) compares the mean flow found analytically from
Eq.~(\ref{eq:vorticity_potential}) with the mean flow from
simulation flowing along the x-axis at $y=0$. Parameters are
$r_0=11.31$, $r_c=13.20$, $r_1=20.0$, $\delta_r=0.036$,
$\sigma=0.87$ and $\epsilon_0 = 0.025$.}
\label{fig:amp_mf_vort_eps_exp1750}
\end{figure}

It is illustrative to compare the analytical results of
Section~\ref{section:analytical development} with the results of
simulation. Figure~\ref{fig:amp_mf_vort_eps_exp1750}a displays
$A^2(r)$ for the case $\epsilon_0=0.025$, as determined by
Eq.~(\ref{eq:amp_nondim}). A significant nonadiabaticity is
present for this case as shown by the deviation of $A^2(r)$ from
$\epsilon(r)/g_o$. For the ramped domain used in simulation the
distance $r_c(\epsilon_0)-r_0$ is smaller than in the larger
domain with a more shallow ramp used in experiment. This results
in the presence of more nonadiabaticity in the simulations when
compared to experimental results at the same control parameter.
This is beneficial because this allows the exploration of highly
nonadiabatic situations without having to perform the task of
simulating near the convective threshold, which becomes
computationally difficult because of the diverging time scales.

A comparison between theory and simulation of the vertical
vorticity and the resulting mean flow is shown in
Fig.~\ref{fig:amp_mf_vort_eps_exp1750}b and c. The theoretical
predictions are based on the amplitude variation caused when
straight parallel convection rolls encounter a radial ramp in
plate separation as discussed in Section~\ref{section:analytical
development}. For both the vertical vorticity and the mean flow
the comparison is made in the absence of any adjustable
parameters. For the vertical vorticity calculated in simulation an
angular average, weighted by $\sin 2 \theta$, is used for the
comparison. The agreement between theory and simulation is quite
good. This illustrates quantitatively that the major source of
vertical vorticity and mean flow is indeed the variation in the
convective amplitude caused by the radial ramp in plate
separation. Over the bulk of the domain the mean flow is negative
and very small in magnitude with a maximum value of $U_x(y=0)=-7.0
\times 10^{-4}$ and by Eq.~(\ref{eq:delta_k}) the wavenumber
variation would be extremely small in agreement with the near
constant bulk wavenumbers found in simulation.
%%%%%%%%%%%%%%%%%%%%%%%%%%%%%%%%%%%%%%%%%%%%%%%%%%%%%%%%%%%%%%%%%%%%%%%%%%%%%
%%%%%%%%%%%%%%%%%%%%%%%%%%%%%%%%%%%%%%%%%%%%%%%%%%%%%%%%%%%%%%%%%%%%%%%%%%%%%

\begin{figure}[ht]
\begin{center}
\includegraphics[width=2.5in]{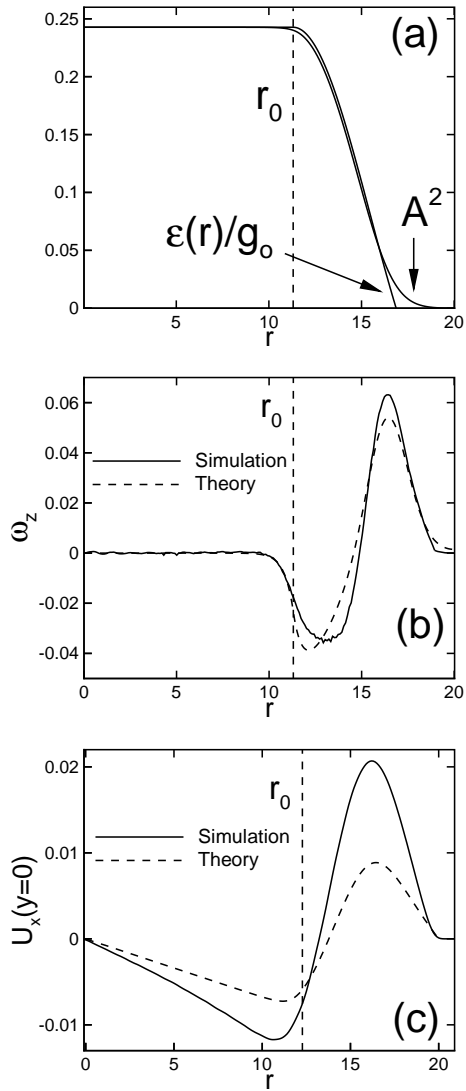}
\end{center}
\caption{Panel~(a) shows the solution of Eq.~(\ref{eq:amp_nondim})
plotted as $A^2(r)$, shown for comparison is $\epsilon(r)/g_o$.
Panel~b compares the vertical vorticity found analytically from
Eq.~(\ref{eq:amp_nondim}) with an angular average, weighted by
$\sin 2 \theta$, of the vertical vorticity from simulation.
Panel~(c) compares the mean flow found analytically from
Eq.~(\ref{eq:vorticity_potential}) with the mean flow from
simulation flowing along the x-axis at $y=0$. Parameters are
$r_0=11.31$, $r_c=13.20$, $r_1=20.0$, $\delta_r=0.036$,
$\sigma=0.87$ and $\epsilon_0 = 0.171$.}
\label{fig:amp_mf_vort_eps_exp2000}
\end{figure}

A similar comparison between theory and simulation is made in
Fig.~\ref{fig:amp_mf_vort_eps_exp2000} for $\epsilon_0=0.171$. As
shown in Fig.~\ref{fig:amp_mf_vort_eps_exp2000}a $A^2(r)$ is much
more able to follow the ramp, $\epsilon(r)/g_o$, and exhibits very
little nonadiabaticity except for the kink near $r_c$. This
results in a much stronger negative vertical vorticity in the bulk
which in turn yields a larger negative mean flow as shown if
Fig.~\ref{fig:amp_mf_vort_eps_exp2000}b and c. The agreement
between theory and simulation for the vertical vorticity is still
quite good. The discrepancy in the mean flow comparison may be due
to the fact that as $\epsilon_0$ increases other mean flow sources
such as roll curvature, see Fig.~\ref{fig:tp}b, become important.
%%%%%%%%%%%%%%%%%%%%%%%%%%%%%%%%%%%%%%%%%%%%%%%%%%%%%%%%%%%%%%%%%%%%%%%%%%%%%
%%%%%%%%%%%%%%%%%%%%%%%%%%%%%%%%%%%%%%%%%%%%%%%%%%%%%%%%%%%%%%%%%%%%%%%%%%%%%

\begin{figure}[ht]
\begin{center}
\includegraphics[width=2.5in]{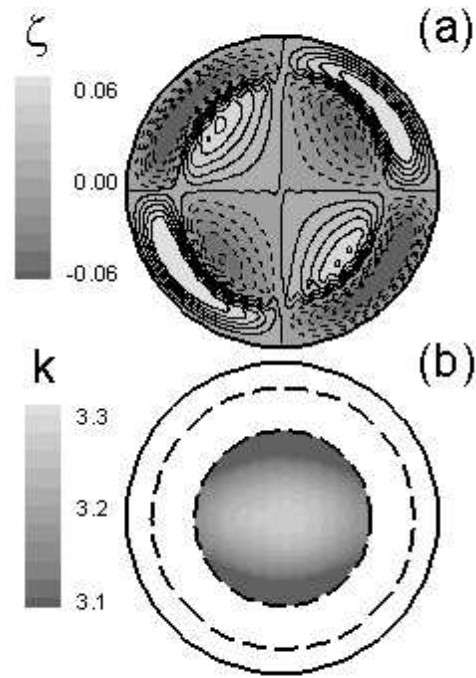}
\end{center}
\caption{Contours of the vorticity potential $\zeta$, panel~(a),
(light indicates counterclockwise rotation drawn with solid
contours and dark indicates clockwise rotation drawn with dashed
contours) and the corresponding local wavenumber distribution,
$k$, panel~(b). The magnitude of the mean flow is approximately
$2\%$ of the magnitude of the velocity field,
$|\vec{U}_s|/|\vec{u}| \approx 0.02$. Local wavenumber
distributions are shown only in the bulk, $r \le r_0$. The inner
dotted circle indicates where the ramp begins, $r_0$, and the
outer dotted circle indicates where the convection layer becomes
critical, $r_c$. Simulation parameters, $r_0=11.31$, $r_1=20$,
$\delta_r=0.036$, $\sigma=0.87$ and $\epsilon_0=0.171$
($R_0=2000$).} \label{fig:mf_wn}
\end{figure}

Figure~\ref{fig:mf_wn} illustrates the octupole structure in the
vorticity potential in panel~(a) and the roll compression
occurring in the bulk by plotting contours  of the local
wavenumber in panel~(b) for $\epsilon_0 = 0.171$. As illustrated
in panel~(a) the mean flow has significant structure over the
ramped region as well as extending into the subcritical region of
the layer, $r>r_c$. It has also been suggested that the mean flow
extends into a subcritical region in related experiments
implementing ``finned" boundaries~\cite{pocheau:1997}.

The vorticity potential displays an octupole structure containing
a pair of counter-rotating vortices in each quadrant. The inner
quadrupole is localized around $r_0$ where gradients in the
amplitude of convection occur as the ramp in plate separation
begins. The direction of rotation of the inner quadrupole causes a
focussing of the mean flow into the bulk region of the domain and
is responsible for the larger wavenumbers found as $\epsilon_0$ is
increased as shown by the ($\circ$) curve in Fig.~\ref{fig:wnall}.
%%%%%%%%%%%%%%%%%%%%%%%%%%%%%%%%%%%%%%%%%%%%%%%%%%%%%%%%%%%%%%%%%%%%%%%%%%%%%
%%%%%%%%%%%%%%%%%%%%%%%%%%%%%%%%%%%%%%%%%%%%%%%%%%%%%%%%%%%%%%%%%%%%%%%%%%%%%

\begin{figure}[tbh]
\begin{center}
\includegraphics[width=2.5in]{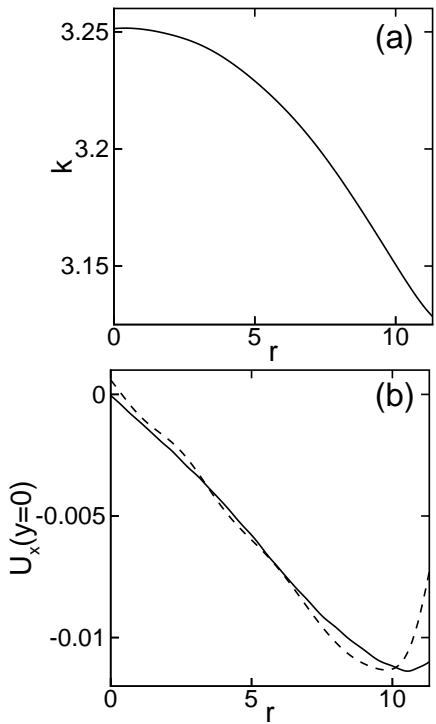}
\end{center}
\caption{Panel~(a), the variation in the local wavenumber along
the positive x-axis, or equivalently $k(r)$ at $\theta=0$.
Panel~(b), a comparison of the mean flow from simulation (solid
line) with the predicted value calculated from
Eq.~(\ref{eq:phase}) using the wavenumber variation from
panel~(a). Simulation parameters, $r_0=11.31$, $r_1=20$,
$\delta_r=0.036$, $\sigma=0.87$ and $\epsilon_0=0.171$
($R_0=2000$).} \label{fig:wn_mf_compare2000}
\end{figure}

To make the connection between mean flow and wavenumber
quantitative Eq.~(\ref{eq:phase}) is applied to the simulation
results in the form,
\begin{equation} \label{eq:phase_new}
U=D_\parallel k_c^{-1}\partial_x k.
\end{equation}
Figure~\ref{fig:wn_mf_compare2000}a illustrates the wavenumber
variation, $k(r)$ for $r \le r_0$, found in simulation by simply
measuring the distance between roll boundaries and makes evident
the roll compression, $k(r=0)>k(r_0)$.
Figure~\ref{fig:wn_mf_compare2000}b compares the mean flow
calculated from simulation with the predicted value of the mean
flow required to produce the wavenumber variation shown in
Fig.~\ref{fig:wn_mf_compare2000}a using Eq.~(\ref{eq:phase_new}).
The agreement is good and the discrepancy near $r_0$, which is
contained within one roll wavelength from where the ramp begins,
is expected because the influence of the ramp was not included in
Eq.~(\ref{eq:phase}). This illustrates quantitatively that the
mean flow compresses the rolls in the bulk of the domain.
%%%%%%%%%%%%%%%%%%%%%%%%%%%%%%%%%%%%%%%%%%%%%%%%%%%%%%%%%%%%%%%%%%%%%%%%%%%%%
%%%%%%%%%%%%%%%%%%%%%%%%%%%%%%%%%%%%%%%%%%%%%%%%%%%%%%%%%%%%%%%%%%%%%%%%%%%%%

As mentioned earlier, the mean flow vanishes as $\epsilon_0$
approaches critical whereas the large scale counterflow is present
for all $\epsilon_0$ and therefore could play a role near
threshold in the determination of the final convection pattern. In
order to gain further insight into this possibility a radial ramp
was constructed that did not drive a large scale counterflow. This
was accomplished by setting the temperature of the ramped surface,
$T_b(r)$, to the value of the linear conduction profile at that
height, $T_b(r)=h(r)$. This ramp, therefore, does not bend the
isotherms which is the source of the large scale counterflow. The
wavenumber variation for these simulations, see curve labelled
with ($\triangleright$) in Fig.~\ref{fig:wnall}, does not differ
strongly from the simulations with a ramp producing large scale
counterflow, see the curve labelled with ($\circ$). The similarity
in wavenumber results is strongest for small $\epsilon_0$
suggesting that the large scale counterflow is not responsible for
the shift of the critical wavenumber to smaller values as seen in
experiment.

To study the large scale counterflow further two-dimensional
simulations were also performed, corresponding to a vertical slice
of the domain considered thus far, and in addition to a more
spatially extended domain as used in experiment. In two dimensions
the mean flow is absent however the large scale counterflow
persists. As shown by the ($\diamond$) and ($\square$) curves in
Fig.~\ref{fig:wnall} the wavenumbers measured in the
two-dimensional simulations are not compressed to the same extent
as $\epsilon_0$ increases as in the three-dimensional simulations
with both mean flow and large scale counterflow present. As
expected, the wavenumbers found in the two-dimensional simulations
are also independent of aspect ratio. Additionally, for small
$\epsilon_0$ the wavenumber found in simulation does not deviate
markedly from its critical value suggesting that the large scale
counterflow is not responsible for the wavenumber shift observed
in experiment near threshold regardless of the spatial extent of
the domain.

We also investigated the possibility that the vertical large scale
counterflow could bifurcate into a horizontal flow similar to the
mean flow in the presence of a slight spatial asymmetry. This was
accomplished by giving the ramped domain used in the full
three-dimensional simulations an eccentricity of $e \approx 0.8$
for a variety of ramps $0.036 \le \delta_r \le 0.25$ and
simulating over a range of subcritical and supercritical
conditions. For all of the scenarios tested the large scale
counterflow remained vertical and did not undergo any significant
changes.

Lastly, the possibility of wavenumber pinning was studied by
varying the aspect ratio in increments of less than half of a roll
width for both the two- and three-dimensional domains. In all of
the scenarios tested the final pattern wavenumbers were not
appreciably affected by these small changes in aspect ratio.
\section{Conclusion}
\label{section:conclusion} We have analytically and numerically
investigated pattern formation in a cylindrical convection layer
with a radial ramp in plate separation. In particular, we have
studied quantitatively the effects of two large scale flows; large
scale counterflow and mean flow. These large scale flows are
important theoretically yet are extremely difficult to measure
experimentally.

Our results suggest that the mean flow plays an important role in
the observed pattern wavenumber and is generated in a novel way,
by the spatial variation of $|A|^2$ driven by the variation of
$\epsilon$ rather than the more usual variations in roll curvature
and wavenumber. The mean flow sources are quantified analytically
and agreement is found with numerical results. The geometric
structure and magnitude of the mean flow is used to explain
quantitatively the wavenumber variation found in the simulations.

The large scale counterflow is investigated numerically and our
results indicate a small roll compression effect away from
threshold. In particular, although the large scale counterflow is
present at and near threshold it does not appear responsible for
the dramatic wavenumber shift to values less than critical as seen
in experiment.

Although it is too expensive computationally to simulate the very
large systems used in the experiments we can use our quantitative
understanding of the ramp-generated mean flow, validated by the
simulations at smaller aspect ratio, to extrapolate our results to
these larger systems. Furthermore our two-dimensional simulations
in sizes equal to the experimental ones allow us to estimate the
effect of the large scale counterflow on the wavenumber
distribution. Despite these exhaustive efforts, we are unable to
reproduce the large shift to smaller wavenumbers observed near
threshold in the experiments, and the physical origin of these
results remains a mystery.
\section*{ACKNOWLEDGEMENTS} We are
grateful to H. S. Greenside and G. Ahlers for helpful discussions.
This research was supported by the U.S.~Department of Energy,
Grant DE-FT02-98ER14892, and the Mathematical, Information, and
Computational Sciences Division subprogram of the Office of
Advanced Scientific Computing Research, U.S.~Department of Energy,
under Contract W-31-109-Eng-38. We also acknowledge the Caltech
Center for Advanced Computing Research and the North Carolina
Supercomputing Center.

%\bibliography{all}
\end{document}